\documentclass[]{spie}  %>>> use for US letter paper

\usepackage[]{graphicx}
\usepackage{epstopdf}
\title{Phase space monitoring of Exciton-polariton multistability}

\author{Yoan L\'eger
\skiplinehalf
Foton Laboratory, Universit\'e Europ\'eenne de Bretagne, CNRS-INSA-UR1, INSA de Rennes, F-35708 Rennes, France.
}

\authorinfo{Further author information: (Send correspondence to Y.L.)\\Y.L.: E-mail: yoan.leger@insa-rennes.fr, Telephone: 33 2 23 23 88 61}
\begin{document}
  \maketitle

%%%%%%%%%%%%%%%%%%%%%%%%%%%%%%%%%%%%%%%%%%%%%%%%%%%%%%%%%%%%%
\begin{abstract}
Dynamics of exciton-polariton multistability is theoretically investigated. Phase portraits are used as a tool to enlighten the microscopic phenomena which influence spin multistability of a confined polariton field as well as ultrafast reversible spin switching. The formation of  a non-radiative reservoir, due to polariton pairing into biexcitons is found to play the lead role in the previously reported spin switching experiments. Ways to tailor this reservoir formation are discussed in order to obtain optimal spin switching reliability.
\end{abstract}

%>>>> Include a list of keywords after the abstract


%%%%%%%%%%%%%%%%%%%%%%%%%%%%%%%%%%%%%%%%%%%%%%%%%%%%%%%%%%%%%
\section{INTRODUCTION}
\label{sec:intro}  % \label{} allows reference to this section
For more than a decade, exciton-polariton physics has boomed, both on theoretical and experimental sides, in applied and fundamental science. In 2012, polariton-related publications have crossed the line of 2500 citations per year. This success originates in the very nature of these quasi-particles. Polaritons are formed in VCSEL-like semiconductor microcavities of high optical and material quality. The magnified interaction between confined photons and quantum well excitons leads to the anticrossing of the light and matter resonances, producing these unique quasi-particles. Polaritons take the best of both their components. From the photon, they inherit the very light mass (five order of magnitude below the free electron mass due to the steep optical cavity dispersion) and the coherence. From the exciton, they inherit the richness of spinor interactions. But the main interest of polaritons is their bosonic nature. The demonstration of Bose Einstein condensation of polaritons\cite{kasprzak_bose-einstein_2006} triggered a huge series of seminal works taking advantage of both polaritons' bosonic character or interactions: superfluidity,\cite{amo_superfluidity_2009} half-quantized vortices,\cite{lagoudakis_observation_2009} solitons,\cite{amo_polariton_2011, nardin_hydrodynamic_2011} spin-Hall effect,\cite{leyder_observation_2007} spin multistability\cite{gippius_polarization_2007, paraiso_multistability_2010, vishnevsky_multistability_2012}... Polaritons have also progressively entered applied Physics; they have been proposed as optical transistors,\cite{shelykh_optically_2010, gao_polariton_2012, ballarini_all-optical_2013} all-optical spin memories \cite{cerna_ultrafast_2013}, switches\cite{amoa_exciton-polariton_2010} or even for neuronal optical circuitry.\cite{liew_optical_2008}

In this work, polariton multistability is scrutinized by monitoring phase portraits of the polariton system. The aim is here to illustrate and deepen the understanding of the polariton spin dynamics and in particular the reliability of spin switching experiments realized by Cerna et al.\cite{cerna_ultrafast_2013} The model used in this article notably includes the formation of a non-radiative reservoir due to the pairing of polaritons into biexcitons.\cite{wouters_influence_2013} As experimentally demonstrated, this reservoir plays a crucial role in spin multistability, not only because of the nonlinear losses resulting in its formation but also because of its subsequent interactions with polaritons. Here we demonstrate through phase space analysis that the reservoir formation can be tailored to optimize spin switching processes in polariton systems.
%%%%%%%%%%%%%%%%%%%%%%%%%%%%%%%%%%%%%%%%%%%%%%%%%%%%%%%%%%%%%
\section{From bistability to multistability}

%%-----------------------------------------------------------
\subsection{The exciton-polariton system}
\label{sec:polSys}

Exciton-polaritons are formed due to the strong coupling between the confined photon modes and the quantum well excitons of a semiconductor microcavity. Two polariton modes are produced due to this anticrossing. They are called lower and upper polariton modes. In the present work we will focus on lower polariton modes, which show the simplest behavior as they are the ground state of the system. The lifetime-limited coherence of polaritons ranges between a few picoseconds to a hundred of picoseconds.\cite{nelsen_coherent_2012}   Nonlinearities due to polariton-polariton interactions are particularly strong compared to other weakly-coupled systems,\cite{liu_ultra-small_2010}. They are also stronger in materials with weak exciton binding energy such as InGaAs and GaAs and thus large Bohr radii, which unfortunately penalizes large band-gap semiconductors where room temperature lasing can occur, such as ZnO and GaN.\cite{christmann_room_2008} Due to their 2D character, polaritons feature a pseudo-spin 1/2, equivalent to the ellipticity of light in the emission direction of the cavity. In the following we will mention the two polariton spin projection as $\sigma_+$ and $\sigma_-$ states. The polariton-polariton interaction is highly spin dependent due to Pauli exclusion and excitonic complex formation.\cite{kwong_third-order_2001, vladimirova_polariton-polariton_2010-1, takemura_polariton_2013} The copolarized polariton interaction (which coefficient is called $\alpha_1$ in the following) is repulsive while the co-polarized polariton interaction ($\alpha_2$) is usually considered attractive and much smaller than the copolarized interaction.

As detailed below, phenomena such as bistability and multistability highly depend on the linewidth of the polariton resonance. Bistability could be observed in a planar microcavity with Q-factor about 5'000 ten years ago.\cite{baas_optical_2004} However, spinor multistability is more difficult to observe because of the competition between the polarized modes. It was first observed in polariton mesas where the particles are confined in the 3 directions, limiting photonic-disorder-induced linewidth broadening (Q-factor about 10'000).\cite{paraiso_multistability_2010} The full confinement of polaritons greatly helps to obtain sharp and well defined multistability hystereses as it has been seen experimentally.

%%-----------------------------------------------------------
\subsection{Bistability}
To understand polariton bistability we can first restrict the system to the lower polariton modes and spinless particles. One can observe polariton bistability by different ways: In a wedged microcavity, where the cavity resonance depends on the spatial position, one can change continuously the detuning between the polariton resonance and the laser line by scanning the excitation spot position of a cw injection laser. This is the first experimental demonstration of this phenomenon.\cite{baas_optical_2004} Reciprocally, one could also scan the wavelength of a tunable cw injection laser at a given place on the sample. The last possibility is to detune the injection laser above the polariton resonance and vary the injection power. The latter protocol has been used to observe multistability of confined polaritons.\cite{paraiso_multistability_2010} We will focus on this one in the present article.

The master equation of the bistable polariton system is a simple nonlinear Shr\"{o}dinger equation:
\begin{equation}
\frac{d\psi}{dt}=(-\frac{\gamma}{2}-i(\omega_0+\alpha_1|\psi|^2))\psi+Fe^{-i\omega_Ft} \label{eq:bista}
\end{equation}
where $\psi$ is the lower polariton field, $\gamma$, its linewidth and $\omega_0$ its resonance energy. As mentioned above, $\alpha_1$ refers to the polariton-polariton interaction coefficient. The cw injection laser field is defined by its amplitude and frequency, $F$ and $\omega_F$ respectively.

Stationary solutions of this nonlinear equation are obtained by first solving the intensity relation, where $I_F=F^\star F$, $I=\psi^\star \psi$ and $\delta=\omega_F-\omega_0$:
\begin{equation}
I_F=\left[(\frac{\gamma}{2})^2+(\alpha_1I-\delta)^2\right]I.\label{eq:Ipoly}
\end{equation}
The non-bijective character of $I_F$ as a function of $I$ when $\delta>\sqrt{3}\gamma/2$ defines the bistability condition: The laser detuning should be much larger than the resonance linewidth for bistability to occur. An example of bistability hysteresis is provided in Fig.1 a with parameters close to standard polariton systems. One can then make the polariton field explicit:
\begin{equation}
\psi=\frac{F}{\gamma/2+i(\alpha_1I-\delta)},\label{eq:Ipoly}
\end{equation}

A convenient way to map the bistability phenomenon consists in the use of phase portraits.\cite{evans_advances_2004} They show an ensemble of trajectories of the nonlinear system for different initial conditions in phase space. In the present case, where $F$ is fixed, phases can be either $(\Re(\psi),\Im(\psi))$ or $(|\psi|,\arg(\psi))$. We will use the latter representation since it provides more straightforwardly information on phase and polarization relations between polariton fields and injection fields.  We will use the injection laser phase as the phase reference for the polariton fields.

\begin{figure}
   \begin{center}
   \begin{tabular}{c}
   \includegraphics[width=12cm]{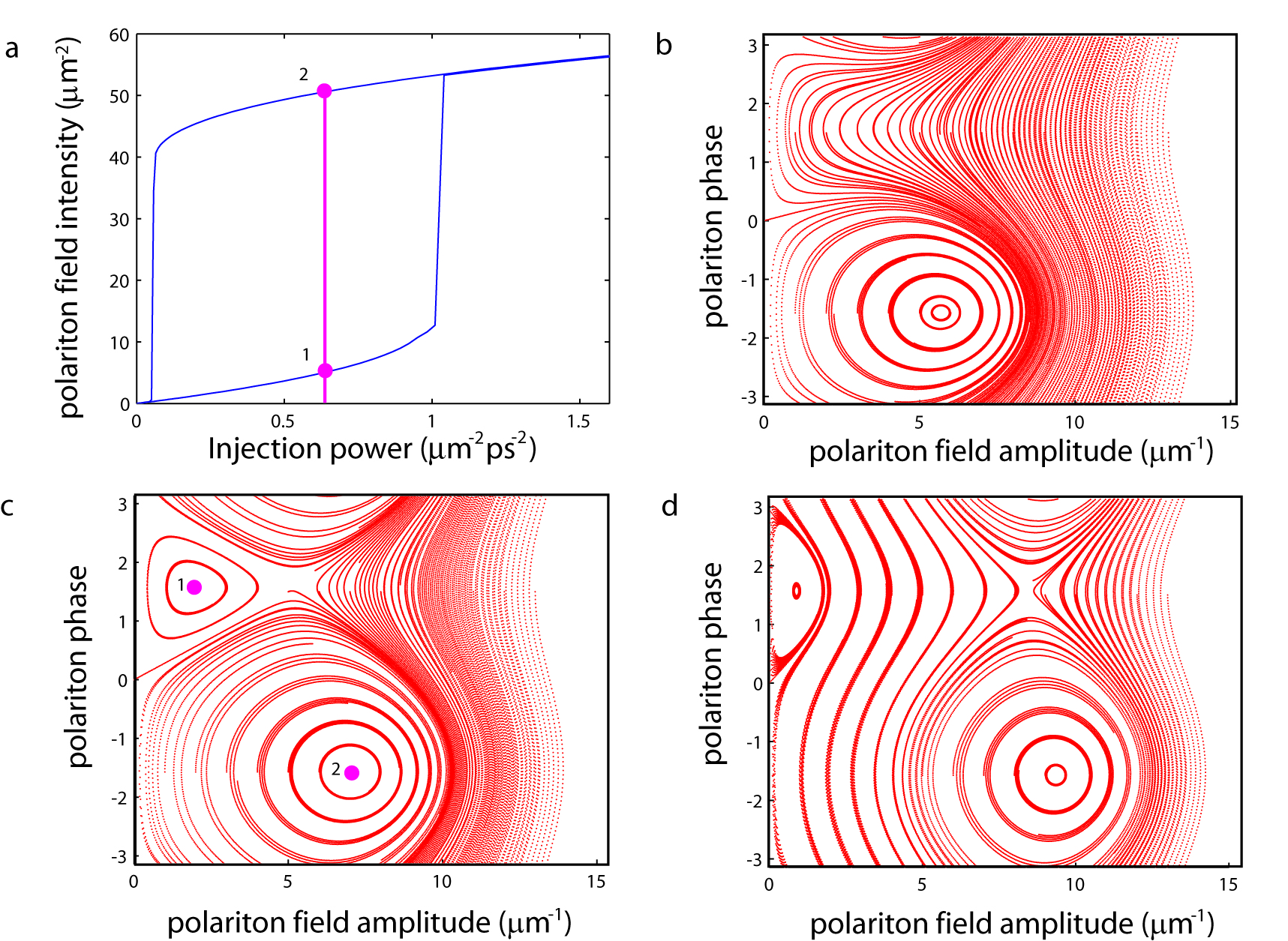}
   \end{tabular}
   \end{center}
   \caption
{ \label{fig:1}
(a) Power dependent bistability hysteresis of a polariton resonance ($\delta=0.4$ meV, $\alpha_1=0.01$ meV.$\mu$m$^2$, $\gamma=0.08$ meV). (b,c,d) Phase portraits of the polariton bistable system for zero linewidth, injection power $I_F=0.7\mu m^{-2}ps^{-2}$, zero pump phase and different laser detunings: b, $\delta=0.2$ meV; c, $\delta=0.4$ meV; d, $\delta=0.8$ meV.}
\end{figure}

To sketch qualitatively the phase-space landscape of a bistable system, the damping term $\gamma$ is generally first set to zero.\cite{evans_advances_2004} This leads to a set of closed orbit plots, similar to a contour plot of the phase-plane landscape and which reveals the stability solutions of the system (so-called fixed points). Fig.1 b,c and d provide such phase portraits of the polariton bistable system for three different detunings $\delta$. For Fig.1 c, the two fixed points are labeled $1$ and $2$ and correspond to the two solutions of Fig.1 a for an injection power $I_F=0.7\mu m^{-2}ps^{-2}$. When the detuning is very small (Fig.1 b), such an injection power is enough to place the system out of the hysteresis on the high intensity side. Only the high-intensity fixed point can be seen in the phase portrait. In this case, the polariton resonance jumped above the laser resonance due to the nonlinear blueshift. The polariton field is thus $-\pi/2$ shifted compared to the injection laser, such as for any driven oscillator.
As the laser detuning is increased, the system enters the bistability area (Fig.1 c an d). The low intensity fixed point appears on the phase portraits. Its phase is $\pi/2$, showing that the polariton resonance is below the laser line. We also observe that the two fixed points are progressively pushed apart in the amplitude direction. The evolution of the phase space landscape as a function of detuning will be determinant to understand multistability.

When damping is introduced, the system trajectories relax to one or the other of the fixed points just like a marble oscillating in a bowl, as it can be seen in Fig.2. But more interestingly,  the increase of the damping coefficient also modifies the phase-space landscape. It makes the high intensity fixed point less and less attractive. When the bistability condition $\delta>\sqrt{3}\gamma/2$ is not fulfilled anymore (Fig.2 c), the high intensity fixed point disappears. The reader will also note that the position of the fixed points changes when varying the damping coefficient. This can be attributed to the variation of the denominator argument with damping in eq.(\ref{eq:Ipoly}). The influence of damping on the phase space landscape is also of prime importance in the multistability mechanism as discussed below.

\begin{figure}
   \begin{center}
   \begin{tabular}{c}
   \includegraphics[width=15cm]{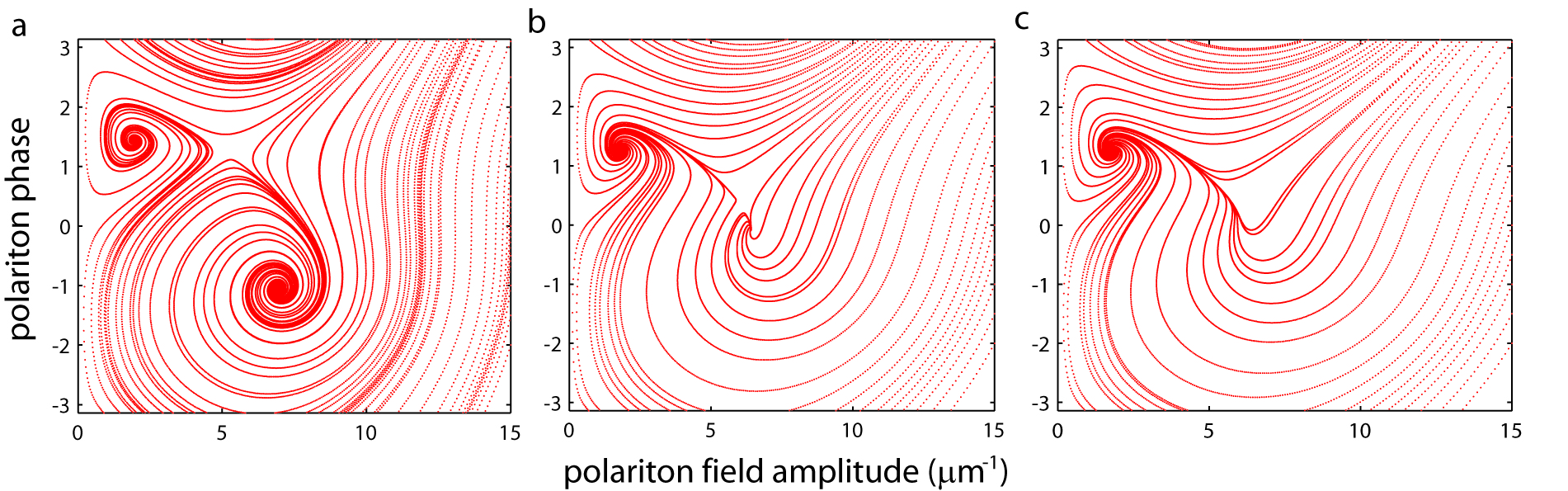}
   \end{tabular}
   \end{center}
   \caption
{ \label{fig:2}
Phase portraits of the polariton bistable system for different polariton linewidths: a, $\gamma=0.1$ meV; b, $\gamma=0.22$ meV; c, $\gamma=0.24$ meV. In the latter case, bistability condition $\delta\geq\sqrt{3}\gamma/2$ is not fulfilled.}
   \end{figure}
%%-----------------------------------------------------------
\subsection{Spin dependent nonlinearities}
In the previous section, we considered spinless polaritons. When polariton spin is introduced, two degenerate modes, corresponding to the left and right circular polarization states of light ($\sigma_\pm$) participate in the nonlinear dynamics. As mentioned in section \ref{sec:polSys}, the spinor polariton interaction is anisotropic. This ensures a partial decoupling of the bistability hystereses that occur for each polarization mode. This observation allowed Gippius et al. to predict polariton spin multistability:\cite{gippius_polarization_2007} For a given injection condition, each polarization mode of the doublet can be either in a low or high intensity state, authorizing up to four fixed points for the nonlinear two mode system:low intensity for both polarizations, $\sigma_+$ emission, $\sigma_-$ emission, and linearly polarized emission (both polarization in the high intensity state).  However, the experimental demonstration of spin multistability showed that the $\sigma_+$ and $\sigma_-$ hystereses are not as decoupled as expected.\cite{paraiso_multistability_2010} A theoretical reproduction of the experimental multistability behavior is presented in Fig.3 a.
First, one observes that the intensity threshold allowing the modes to jump to the high intensity branch ($I_F=6\mu m^{-2}ps^{-2}$) is the same for both polariton spins. This proves that despite the attractive nature of the counter-polarized spin interaction, the global polariton dynamics result in an effective repulsive interaction for counter-polarized polaritons.
Second, when decreasing the injection power from $6\mu m^{-2}ps^{-2}$ to $2\mu m^{-2}ps^{-2}$, the $\sigma_-$ intensity drop induces an additional jump of the $\sigma_+$ intensity. This can only be explained by a dependence of the damping on the counter-polarized spin intensity.
Third, This additional jump reveals another hysteresis when the injection is increased again (gray arrows on Fig.3 a). For this injection conditions we have tristability of the polariton system.

These three observations bring Wouters et al.\cite{wouters_influence_2013} to propose the following mechanism as the leading coupling process between counter-polarized polaritons (see Fig.3 b):
When two counter-polarized polaritons scatter, they can form a biexcitonic complex. This results in a spin dependent nonlinear damping scheme for polaritons ($\beta|\psi_\mp|^2$ for $\sigma_\pm$ polaritons). the generated biexciton is not coupled to light anymore and participate in the formation of a long-lived non-radiative reservoir containing biexcitons and also probably high-momentum excitons as a result of biexciton dissociation. This particle reservoir still interacts with polaritons through an additional blueshift. Due to the small overlap integral between polaritons and excitons, the reservoir-induced blueshift coefficient $\alpha_R$ is small compared to the polariton-polariton one, $\alpha_1$. Note that the process of complex formation had already been introduced in Kwong et al.\cite{kwong_third-order_2001} to predict the variation of the counter-polarized polariton interaction $\alpha_2$. The formation of the reservoir is further justified by the resonance of the multistability phenomenon when the lower polariton resonance lies precisely at half the energy of the bare biexciton.\cite{wouters_influence_2013} Effects of the thermal generation of a reservoir under resonant excitation has also been discussed by vishnevsky et al. \cite{vishnevsky_multistability_2012}

However, this model requires to take into account the strong-coupling process with Rabi frequency $\Omega$. Indeed, the large difference between polariton and reservoir lifetimes (at leat one order of magnitude)can easily lead to the non-physical divergence of the reservoir population and of the reservoir-induced blueshift of the polariton resonance. Considering strongly-coupled photon and exciton resonances ensures convergence since the blueshift of the exciton resonance progressively kills the coupling to the photon resonance and thus to the injection.

With these new ingredients, the analysis of polariton multistability consists in solving the following set of five nonlinear equations:
\begin{eqnarray}
\dot{\chi}_\pm&=&\Big(-\big(\frac{\gamma_x}{2}+\beta|\chi_\mp|^2\big)-i\big(\omega_x+\alpha_1|\chi_\pm|^2+\alpha_2|\chi_\mp|^2+\alpha_Rn_R\big)\Big)\chi_\pm-i\frac{\Omega}{2}\varphi_\pm,\label{eq:multi1}\\
\dot{\varphi}_\pm&=&\big(-\frac{\gamma_c}{2}-i\omega_c\big)\varphi_\pm-i\frac{\varepsilon}{2}\varphi_\mp-i\frac{\Omega}{2}\chi_\pm+F_\pm e^{-i\omega_Ft},\label{eq:multi2}\\
\dot{n}_R&=&-\gamma_Rn_R+\beta|\chi_\mp|^2|\chi_\pm|^2,\label{eq:multi3}
\end{eqnarray}
where $\chi$ is the excitonic polarization field, $\varphi$ the confined photon field, $F$ the injection laser field and $n_R$ the reservoir population. For the sake of completeness, the linear polarization splitting $\varepsilon$, which can be due to the anisotropy of the photonic confinement for example, has also been introduced. Such splitting is usually particularly small (a few tens of $\mu$eV\cite{paraiso_multistability_2010}) and only have small consequences on the polariton multistable behavior. We will omit this term in the present work. Moreover, it is quite clear from the experimental multistability behavior that the counter-polarized polariton interaction $\alpha_2$ only contributes weakly to the phenomenon in comparison with the co-polarized interaction and the effects of the reservoir. We will thus set it to zero as well in the following.

Equations (\ref{eq:multi1}), (\ref{eq:multi2}) and (\ref{eq:multi3}) can be solved numerically to analyze phase portraits of the polariton system including the interaction with the reservoir. In the following we study the influence of various initial conditions on the phase space trajectory of the system. For the sake of simplification, we always initialize the trajectories with lower polariton states with given amplitude and phase, i.e. a coherent combination of photon and exciton fields. This choice makes the calculation be very similar to spin switching experiments, where a femtosecond pulse of desired polarization is used to control the polarization state of a multistable lower polariton population.

\begin{figure}
   \begin{center}
   \begin{tabular}{c}
   \includegraphics[width=6cm]{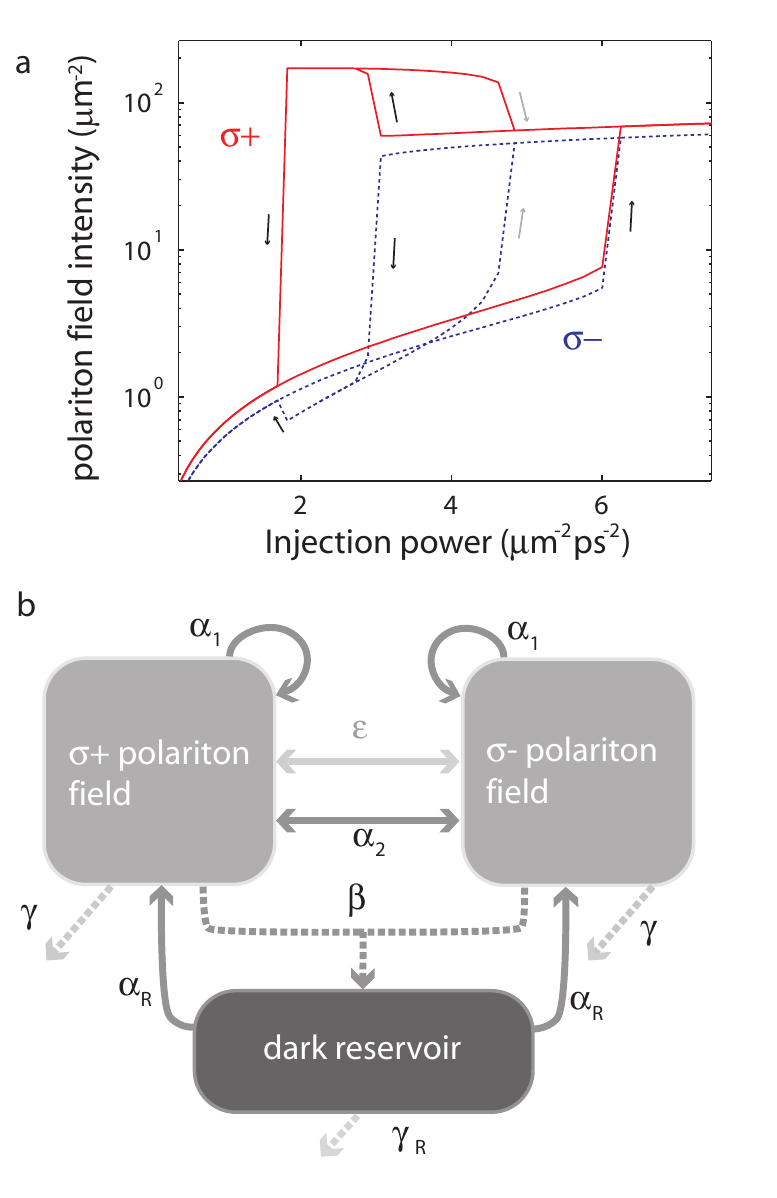}
   \end{tabular}
   \end{center}
   \caption
{ \label{fig:3}
(a) Multistability hysteresis of the spinor polariton system (plain line: $\sigma_+$,dotted line: $\sigma_-$) for a laser polarization slightly favoring $\sigma_+$ injection. (b) Scheme of the polariton system interactions. Energy shifts are plotted with plain arrows and damping with dotted arrows. Nonlinear interactions are plotted in dark gray and linear ones in light gray. }
   \end{figure}

%%%%%%%%%%%%%%%%%%%%%%%%%%%%%%%%%%%%%%%%%%%%%%%%%%%%%%%%%%%%%
\section{Polariton-polariton scattering and reservoir formation} \label{sec:2}

%%-----------------------------------------------------------
\subsection{Effects of cross-polarization nonlinear losses and reservoir formation}

Phase portraits are particularly convenient to analyze the effects of the reservoir on polariton dynamics. We first scrutinize the influence of the initial polariton population on the steady state of the lower polariton polarization doublet. The properties of the cw injection are fixed: it is linearly polarized with an intensity placing the system in the multistability area ($F_+=F_-=1.27\mu m^{-1}ps^{-1}$). The energy detuning between the laser and the lower polariton doublet is $0.4 meV$. Photon, exciton, and reservoir linewidths are the same as in Ref.\cite{cerna_ultrafast_2013} ($\gamma_x=\gamma_c=5\gamma_R=0.05$meV).

It is quite clear from the phase portraits above that the initial phase of the polariton field will influence the steady state of the system since the two fixed points of a bistable system have opposite phases. To enlighten the effect of the initial polariton population, we fix the initial phase of both $\sigma_+$ and $\sigma_-$ polaritons to 1. In other words, the polariton field has the same linear polarization as the cw injection laser. The non-zero relative phase between the cw laser and the polariton field tends to favor relaxation of the system towards the low intensity fixed point.

When the initial polariton field amplitude in each polarization is progressively increased from $5$ to $30\mu m^{-1}$, the phase space trajectory shifts towards the high intensity fixed point (Fig.4 a, b). For initial amplitude larger than $15\mu m^{-1}$ (bold green line), the system does not relax anymore to the low intensity fixed point but to the high intensity one. The trajectories do not feature anymore spiral orbits such as in the low intensity case, showing that the relaxation is faster. These features can be explained by the two effects of the reservoir. First, pairing of polaritons into biexcitons results in an increase of the polariton damping proportional to the counter-polarized polariton population. This explains why the relaxation is faster for the high intensity cases. The relaxation towards the high intensity fixed point is explained by the reservoir-induced blueshift. As it can be seen in Fig.4 c, for low initial amplitude, it takes a while for the system to generate a reservoir population that will blueshift the polaritons significantly. Even if a reservoir-induced blueshift close to the laser detuning can be reached in the $15\mu m^{-1}$ case (bold green line), the system is already in the area of phase space where the relaxation towards the low intensity fixed point is inescapable (light grey area on Fig.4 a and b). The polariton population decays progressively, such as the reservoir. When the initial amplitude is high, the dynamics of the system is more complex. The polariton linewidth is large due to nonlinear losses but decreases rapidly with the polariton population. At the same time the reservoir-induced blueshift (Fig.4 c) increases quite fast, decreasing correspondingly the effective laser detuning and even inverting it. Such as in Fig.1 b (small laser detuning) and Fig.2 c (high damping and bistability condition not fulfilled), there only remains one fixed point for each polariton spin. Since the polariton doublet is sent above the laser line, the new fixed point of the system features a negative phase. the large steady state polariton population is ensured by the balance between the large reservoir-induced blueshift and medium nonlinear losses. The amplitude of the initial polariton field thus strongly influences the steady state of the system. Such large polariton populations are generally easily obtained with ps or fs laser pulses; this enabled the experimental demonstration of polariton-based spin switches and spin memories.

\begin{figure}
   \begin{center}
   \begin{tabular}{c}
   \includegraphics[width=15cm]{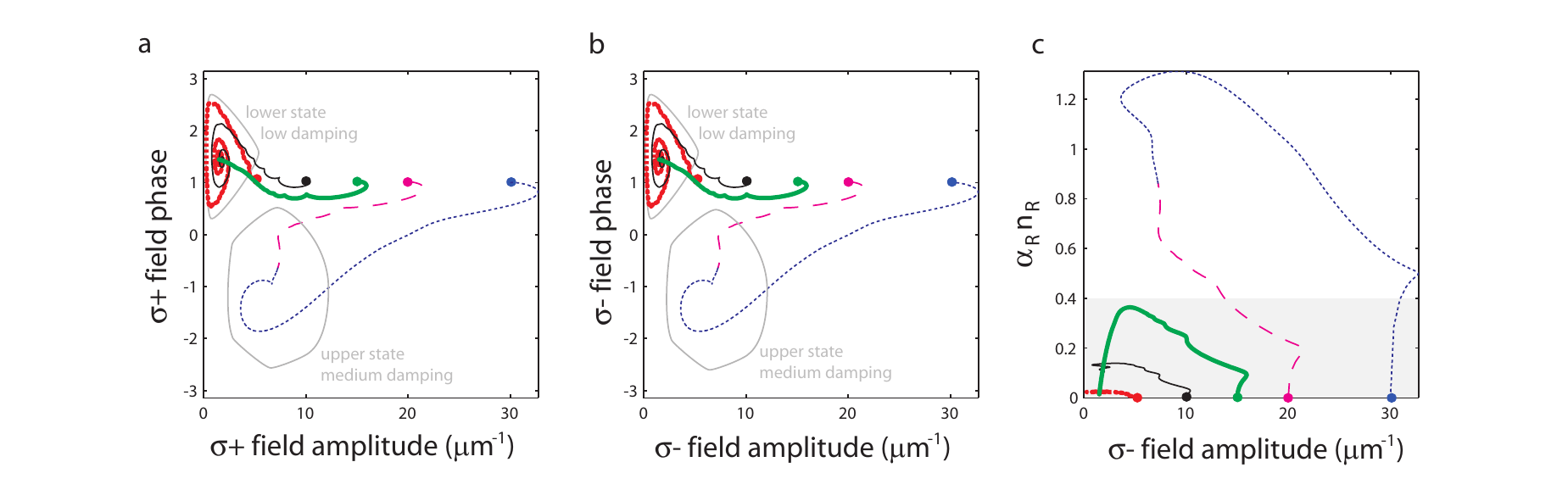}
   \end{tabular}
   \end{center}
   \caption
{ \label{fig:4}
Phase portrait projections of the spinor polariton system for four initial injection conditions: The initial phase is fixed to 1 for all cases but the initial polariton field amplitude is varied (5, 10, 15, 20 and 30 $\mu$m$^{-1}$.). The initial injection polarization is $h$, i.e. similar for $\sigma_+$ and $\sigma_-$ spin projections. as far as upper polaritons are not involved in the phenomenon, access to the amplitude and phase of $\sigma_+$ and $\sigma_-$ lower polaritons and to the reservoir population ($n_R$) gives a comprehensive trajectory of the system in phase space. a (b) $\sigma_{+(-)}$ amplitude/phase plane projection, c Effective reservoir induced blueshift vs $\sigma_{-}$ polariton amplitude. In a and b, stability areas are indicated in light gray. In c, the pump energy detuning is indicated for comparison with the light gray area.}
\end{figure}

%%-----------
\subsection{Tailoring the reservoir formation}
The reservoir formation requires non-negligible polariton populations of both spins. Modifying the trajectory of each polariton spin towards the desired fixed points is thus a way to tailor the reservoir formation and strengthen the achievement of multistable spin switching towards a linearly polarized steady state. The tunability of the initial polariton polarization offers this opportunity. In this section we fix the initial polariton amplitudes of both spins and change the relative phase between them. The initial phase of the $\sigma_+$ polariton field is fixed to 2 while the initial phase of the $\sigma_-$ field is varied between 2, 0 and -2. The phase portraits are presented in Fig.5. The first case is very similar to the dotted line trajectory of Fig.4; the identical phase for both spin populations ensures the convergence towards the same fixed point. Since the initial amplitude is high enough, the polariton fields relax towards a high intensity solution.

Now, when the initial $\sigma_-$ phase is 0, the convergence of $\sigma_-$ polaritons towards a high intensity state is favored. This population enters in the stability area of the high intensity fixed point even before the reservoir-induced blueshift is larger than the laser detuning. At this time, the $\sigma_+$ polariton population is already quite low, such as the nonlinear losses for the $\sigma_-$ polariton population. The latter is thus in a situation pretty close to the case of Fig.1 b (small damping, small laser detuning). The relaxation of $\sigma_-$ polaritons to a high intensity fixed point is compulsory. This causes large additional losses to the $\sigma_+$ polariton field, which relaxation towards the low intensity fixed point is accelerated. Since only one polariton spin population survives in this process, the reservoir formation is blocked and progressively decays to a low value state.

Finally, when the initial $\sigma_-$ phase is set to -2, the $\sigma_-$ polariton field already passed the optimal phase for convergence to the high intensity state. The $\sigma_-$ polariton field feels the presence of the high intensity fixed point but fast bypass it towards low field amplitudes (below 5$\mu$m$^{-1}$). This almost switches off the nonlinear losses for the $\sigma_+$ polariton population. It is now the $\sigma_+$ population that is put in a case close to Fig. 1 b due to the reservoir-induced blueshift.  It relaxes to the high intensity fixed point.

%%-----------
\subsection{Spin switching reliability}
In practical cases however, the laser pulses that initialize the polariton system have a random phase compared to the cw injection laser. From pulse to pulse, the polariton trajectory can thus dramatically change, favoring one or the other of the different fixed points. The only way to strengthen multistable spin switching consists in increasing the initial polariton amplitude. Both the reservoir induced-blueshift and the additional losses will contribute to favor a high intensity fixed point. These contributions are not altered by the phase fluctuations since they are incoherent processes. The initial polarization of the pulse is then the only parameter that determines the system steady state. Switching to $\sigma_+$ and $\sigma_-$ states is easily achieved with circularly polarized one (not discussed here). The most reliable protocol for switching to a linearly polarized polariton state consists in the use of a strong control pulse with linear polarization aligned to the cw laser one. For other linear polarization directions, the system switches from time to time between $\sigma_+$, $\sigma_-$ and linear steady states due to phase fluctuations. Note that in most experiments, integration over several millions of operations can mask the problem of reliability. If a detailed analysis of the polariton intensities can help to discriminate between reliable and non-reliable protocols, polarization interferometry seems to the best way to do it.

\begin{figure}
   \begin{center}
   \begin{tabular}{c}
   \includegraphics[width=15cm]{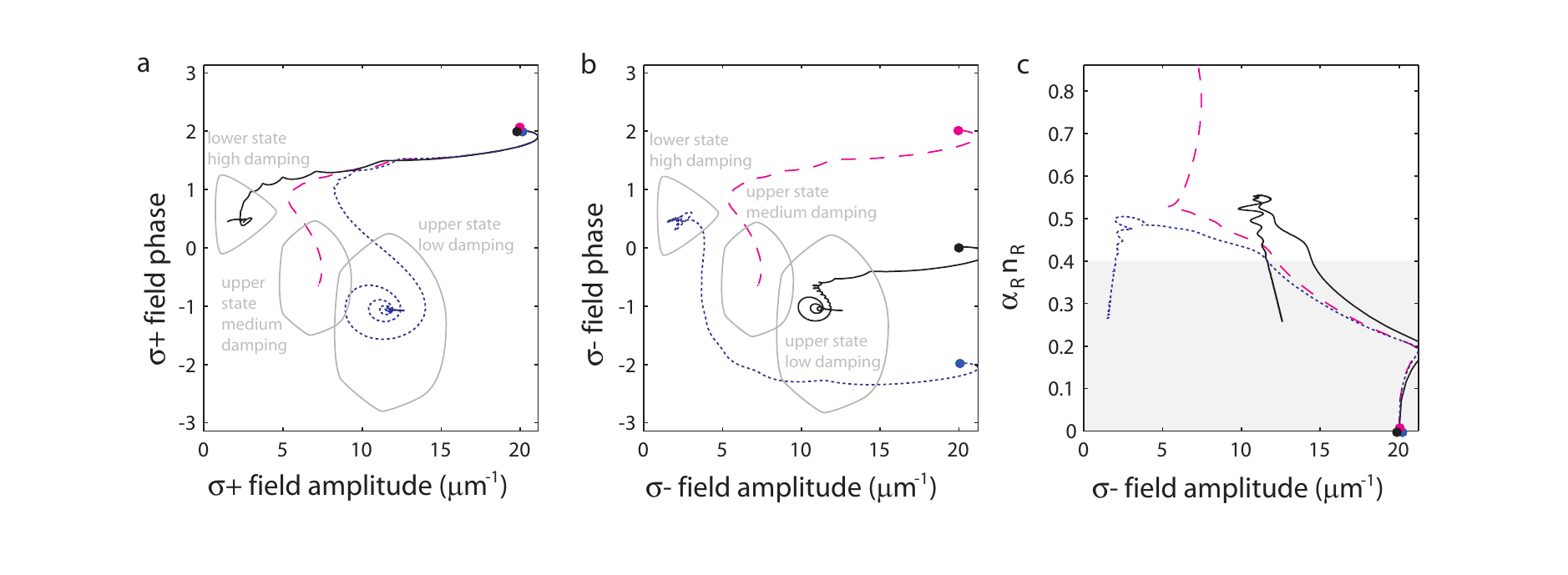}
   \end{tabular}
   \end{center}
   \caption
{ \label{fig:5}
Phase portrait projections (same as Fig.\ref{fig:4}) of the spinor polariton system for three initial polarization conditions: Initial polariton field amplitude is fixed to 20 $\mu$m$^{-1}$; initial $\sigma_+$ polariton phase is fixed to 2 and $\sigma_-$ polariton phase is respectively 2,0 and -2. }
\end{figure}

\section{Conclusion}
The excitonic component of polaritons gives these composite bosons a unique interaction scheme which does not only consist in repulsive/attractive spinor interactions but also in nonlinear losses due to complex formation. To understand the dynamics of polariton spin multistability and spin switching, phase portraits are shown to be particularly useful. We observed the effects of polariton damping and laser-polariton energy detuning on the phase space landscape and linked this analysis to the spin switching phenomenon in a multistable spinor polariton system. The influence of nonlinear losses and reservoir-induced blueshift are found to play a crucial role in the reliability of tristable spin switching.

%\bibliography{bibYL}   %>>>> bibliography data in report.bib

%\bibliographystyle{spiebib}   %>>>> makes bibtex use spiebib.bst

\end{document}